\documentclass[conference]{IEEEtran}
\IEEEoverridecommandlockouts
\usepackage{cite}
\usepackage{amsmath,amssymb,amsfonts}
\usepackage{graphicx}
\usepackage{textcomp}
\usepackage{xcolor}
\usepackage{flushend}
\usepackage{balance}
\usepackage{epsfig}
\usepackage{epstopdf}

\usepackage[linesnumbered,ruled,vlined]{algorithm2e}
\usepackage{float}
\usepackage{subfig}
\usepackage{times}
\usepackage{caption}
\DeclareMathAlphabet{\mathbbold}{U}{bbold}{m}{n}
\usepackage[top=0.71in, bottom=1in, left=0.625in, right=0.625in]{geometry}
\def\BibTeX{{\rm B\kern-.05em{\sc i\kern-.025em b}\kern-.08em
    T\kern-.1667em\lower.7ex\hbox{E}\kern-.125emX}}
\begin{document}

\title{Agentic DDQN-Based Scheduling for Licensed and Unlicensed Band Allocation in Sidelink Networks \vspace{-0.15in} 
\thanks{This work was supported in part by the Academia Sinica (AS) under Grant 235g Postdoctoral Scholar Program, and in part by the National Science and Technology Council (NSTC) of Taiwan under Grant 113-2926-I-001-502-G, 113-2634-F-A49-007, 113-2218-E-A49-027, and 114-2221-E-A49-185-MY3, in part by the National Yang Ming Chiao Tung University (NYCU) under Grant 112UC2N006 and 113UC2N006, and the Ministry of Education (MOE), Taiwan.
Prof. Walid Saad was supported by the U.S. National Science Foundation (NSF) under Grant CNS-2114267. The authors would like to thank Prof. Wan-Jen Huang and Mr. Yen-Ting Liu from the Institute of Communications Engineering (ICE), 
National Sun Yat-sen University (NSYSU) for their valuable assistance and discussions during the early-stage implementation.}
}
\author{\IEEEauthorblockN{Po-Heng Chou$^{1,3}$, Pin-Qi Fu$^{2}$, Walid Saad$^{4}$, and Li-Chun Wang$^{1, 2}$}
\IEEEauthorblockA{
$^{1}$Department of Electronics and Electrical Engineering and $^{2}$Institute of Communications Engineering,\\ National Yang Ming Chiao Tung University (NYCU), Hsinchu 30010, Taiwan\\
$^{3}$Research Center for Information Technology Innovation (CITI), Academia Sinica (AS), Taipei 11529, Taiwan\\
$^{4}$Bradley Department of Electrical and Computer Engineering (ECE), Virginia Tech (VT), Alexandria, VA 22305, USA\\
E-mails: d00942015@ntu.edu.tw, amaza.ee12@nycu.edu.tw, walids@vt.edu, wang@nycu.edu.tw}
\vspace{-0.45in}
}

\maketitle

\begin{abstract}
In this paper, we present an agentic double deep Q-network (DDQN) scheduler for licensed/unlicensed band allocation in New Radio (NR) sidelink (SL) networks. Beyond conventional reward-seeking reinforcement learning (RL), the agent perceives and reasons over a multi-dimensional context that jointly captures queueing delay, link quality, coexistence intensity, and switching stability. A capacity-aware, quality of service (QoS)-constrained reward aligns the agent with goal-oriented scheduling rather than static thresholding. Under constrained bandwidth, the proposed design reduces blocking by up to 87.5\% versus threshold policies while preserving throughput, highlighting the value of context-driven decisions in coexistence-limited NR SL networks. The proposed scheduler is an embodied agent (E-agent) tailored for task-specific, resource-efficient operation at the network edge.
\end{abstract}


\section{Introduction}

Device-to-device (D2D) communications were introduced in 3GPP LTE Release 12~\cite{Release12} to support proximity-based services such as vehicle-to-vehicle (V2V) communication. This concept was extended in 5G New Radio (NR) Release 17~\cite{Release17}, where NR sidelink (SL) networks were standardized for cellular vehicle-to-everything (C-V2X) and other emerging applications. In NR SL networks, resource allocation operates in two modes: a controlled mode, where the base station (gNB) centrally manages allocation, and an autonomous mode, where transmission opportunities are selected by user equipment (UEs) themselves~\cite{TR38875}. Detailed protocol procedures for both modes are specified in 3GPP TR 38.331~\cite{TS38331}.

With the proliferation of Internet of Things (IoT) applications and bandwidth-intensive services, the demand for spectrum resources has outpaced the availability of licensed bands~\cite{Jouhari2023}. Licensed spectrum in 5G NR spans both sub-6 GHz (FR1)~\cite{3GPP381011} and millimeter-wave (mmWave) frequencies (24.25–71 GHz, FR2)~\cite{3GPP381012}. While mmWave bands offer abundant bandwidth, they also suffer from high path loss, susceptibility to blockage, and limited coverage~\cite{TR38901}, necessitating adaptive scheduling strategies. To alleviate congestion, 3GPP Release 16 introduced NR-Unlicensed (NR-U)~\cite{TR38889}, enabling sidelink-unlicensed (SL-U) transmissions in 5 GHz and 6 GHz bands that must coexist with incumbent Wi-Fi systems. Such coexistence, governed by Listen-Before-Talk (LBT)~\cite{TR36889}, often leads to contention, collisions, and degraded quality-of-service (QoS)~\cite{Weerackody2025SLU,Bajracharya2023Bandit}.

These dynamics make resource allocation for NR SL networks uniquely challenging. Unlike traditional cellular scheduling, which can control all channels, SL-U must share the spectrum with uncontrollable Wi-Fi nodes. Prior approaches have attempted threshold-based scheduling~\cite{Chou2019,Chou2020}, rule-driven heuristics, or Markov-based models~\cite{Chou2024ICC}. While valuable, these methods lack adaptability to real-time variations in queueing load, channel fading, and coexistence intensity. Moreover, frequent switching between licensed and unlicensed modes can destabilize performance unless carefully constrained, an aspect often overlooked in prior works.

Reinforcement learning (RL)~\cite{Kaelbling1996} offers a natural solution by enabling agents to adaptively learn from interaction with stochastic environments. Deep Q-networks (DQN)~\cite{Mnih2015DQN,Luong2019Survey} have been widely applied to wireless scheduling, but are prone to Q-value overestimation. Double DQN (DDQN)~\cite{Hasselt2016} addresses this limitation by decoupling action selection from evaluation, achieving more stable convergence. Recent advances in \textit{agentic AI} highlight that agents differ from conventional RL by extending beyond scalar reward optimization toward autonomous, context-driven, and goal-aligned reasoning~\cite{Zhang2025AgentNet}. 
Following this perspective, our DDQN-based scheduler constitutes an embodied agent (E-agent) in~\cite{Zhang2025AgentNet}, by perceiving heterogeneous context, reasoning across layers, and adapting strategies to maintain QoS under constrained spectrum resources.

Motivated by these trends, we develop an agentic DDQN framework that instantiates an E-agent for NR SL networks, jointly allocating licensed and unlicensed spectrum.
Our design integrates queueing delay, channel quality, Wi-Fi coexistence dynamics, and link-switching stability into the state and reward formulation, departing from static heuristics or binary-reward DRL baselines~\cite{Zou2019DRL,Pei2021Q}. 
Unlike conventional RL schedulers that typically optimize a single reward with simplified states (e.g., queue length or binary success), our agent leverages multi-dimensional perception with a capacity-aware, QoS-constrained objective. 
This enables context-driven, goal-oriented reasoning across licensed/unlicensed modes and avoids pathological behaviors such as overusing a single band or oscillating between modes. 

As demonstrated in our experiments, the proposed scheduler achieves up to $87.5\%$ and $82.5\%$ reduction in blocking relative to threshold and random policies, respectively, under constrained bandwidth, while preserving throughput. 
These results highlight the potential of \textit{agentic DDQN}-driven learning, consistent with the RL-oriented definition in~\cite{Zhang2025AgentNet}, to deliver QoS-aware, adaptive, and stable scheduling in coexistence-limited NR SL networks.

\section{Related Work}

RL has become a prominent tool for dynamic spectrum access and adaptive scheduling in unlicensed networks~\cite{Luong2019Survey}. Several studies have applied DQN to address challenges related to NR-U and Wi-Fi coexistence. For example, the authors in~\cite{Zhou2022DRL} proposed an asynchronous multi-channel DQN-based band allocation scheme for NR-U and WiGig systems, focusing on fairness and collision mitigation. The authors in~\cite{Xu2022DRLBeam} developed a DRL-based framework for joint codebook selection and UE scheduling in mmWave NR-U networks, although their approach did not address mode selection or queueing dynamics. Compared with foundation-model-based agents (F-agents) and hybrid agents (H-agents) discussed in~\cite{Zhang2025AgentNet}, our design falls into the E-agent category, emphasizing lightweight, domain-specific adaptation for real-time scheduling.

For SL-U communications, the authors in~\cite{Pei2021Q} applied Q-learning for channel access but relied on simplified state representations and fixed contention parameters, limiting adaptability to dynamic environments. Similarly,~\cite{Yin2020D2DU} introduced a decentralized algorithm based on local interference sensing; however, their approach did not incorporate long-term learning or account for queue dynamics. The authors in~\cite{Bajracharya2023Bandit} developed a multi-objective bandit algorithm to balance fairness and efficiency under LBT constraints, but were limited to NR-U downlink scheduling without dynamic mode-band switching across licensed and unlicensed bands.

DDQN~\cite{Hasselt2016} has been widely applied to wireless systems to improve Q-value stability; however, its application to SL-U scenarios remains largely unexplored. For instance, the authors in~\cite{Zou2019DRL} combined DQN and convex optimization for power control in SL-U networks, but assumed fixed transmission modes and omitted queue-aware scheduling.

In contrast, our study is the first to leverage agentic DDQN for joint mode and unlicensed band selection in queue-aware SL scheduling. Unlike prior works that either use static thresholds~\cite{Chou2019, Chou2020} or binary-reward DRL baselines~\cite{Zou2019DRL, Pei2021Q}, our approach integrates queueing dynamics, throughput–latency tradeoffs, and coexistence constraints under LBT. Moreover, by embedding goal-oriented reasoning into the reward design, the proposed framework aligns with emerging perspectives in agentic DDQN, enabling adaptive and QoS-aware decision-making in highly contention-prone environments.

The main contributions are summarized as follows:
\begin{itemize}
\item \textbf{Agentic DDQN scheduler as an E-agent:} A DDQN agent that perceives multi-dimensional context (queueing, residual spectrum, channel quality, coexistence, switching stability) and reasons toward QoS goals, departing from conventional reward-only RL.

\item \textbf{Unified mode–band action space:} A five-action formulation over CC/SL-L/SL-U and 26/28/5 GHz enables holistic decisions across heterogeneous spectrum options.
\item \textbf{Goal-aligned learning objective:} A capacity-aware reward with an explicit long-term throughput constraint steers policies away from myopic reward chasing and toward stable QoS.
\item \textbf{Coexistence-aware modeling:} LBT/CSMA-CA contention is embedded in the environment, letting the agent adapt SL-U access to incumbent Wi-Fi dynamics.
\end{itemize}

\section{System Model}
\label{section:system_model}

\begin{figure}[h]
    \centering
        \includegraphics[width=0.45\textwidth]{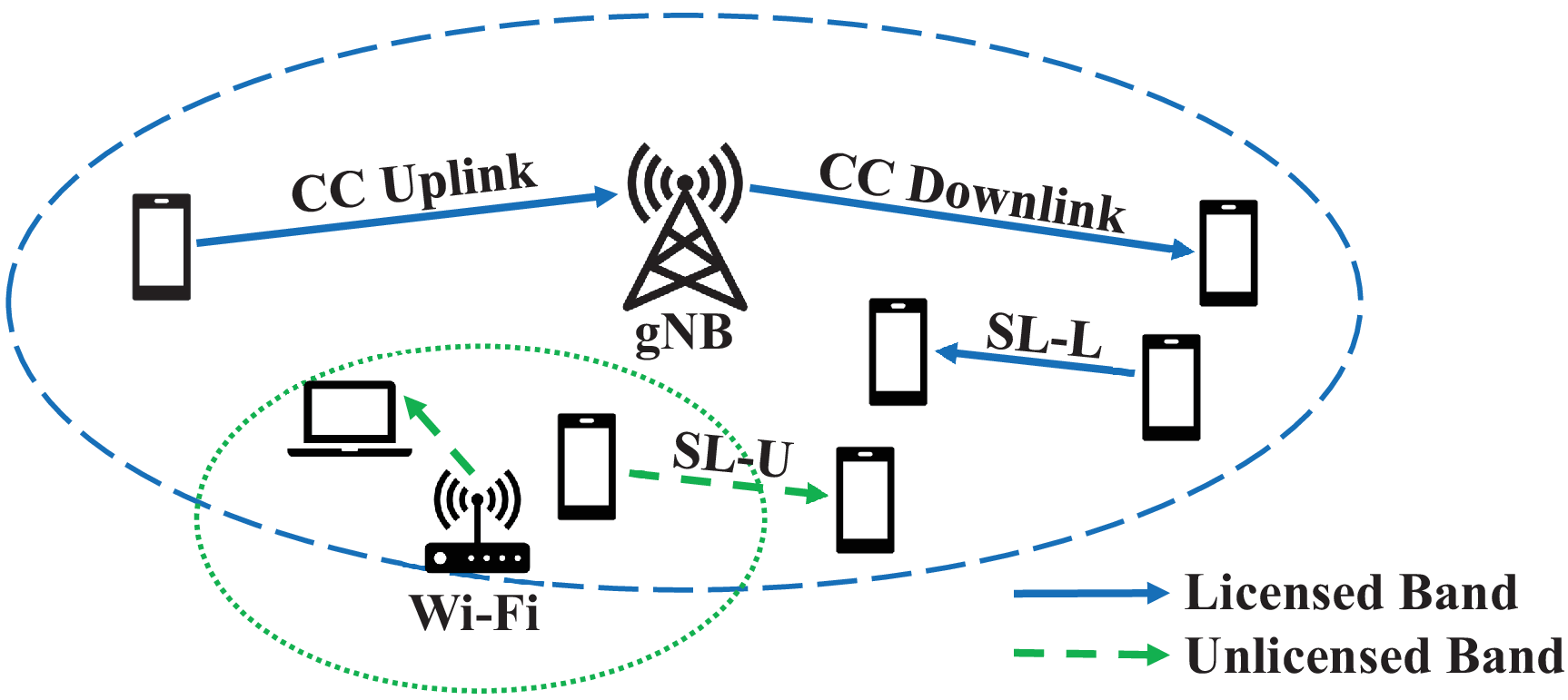}
    \captionsetup{font=small} 
    \caption{Heterogeneous NR SL and Wi-Fi coexistence scenario with licensed (26/28 GHz) and unlicensed (5 GHz) bands, where the gNB allocates CC, SL-L, and SL-U modes.}\label{fig:system_model}
    \vspace{-0.1in}
\end{figure}

We consider a heterogeneous NR SL network where a gNB serves CC and SL users over licensed 26/28 GHz and unlicensed 5 GHz bands, coexisting with Wi-Fi nodes that access the same unlicensed channel via LBT/CSMA-CA, as shown in Fig.~\ref{fig:system_model}. Five 3GPP-compliant transmission options are available: CC and SL-L at 26/28 GHz (FR2) and SL-U at 5 GHz under LBT.

When a new packet arrives at the gNB buffer, the gNB selects one of the following five transmission modes: \begin{enumerate} \item \textbf{CC-28G:} CC via gNB over 28 GHz licensed band; \item \textbf{CC-26G:} CC via gNB over 26 GHz licensed band; \item \textbf{SL-L-28G:} SL transmission over 28 GHz licensed band; \item \textbf{SL-L-26G:} SL transmission over 26 GHz licensed band; \item \textbf{SL-U-5G:} SL transmission over 5 GHz unlicensed band. \end{enumerate}
These transmission options align with 3GPP specifications. CC and SL-L over 26/28 GHz are supported within the FR2 band defined in 3GPP TS 38.901~\cite{TR38901}. SL-U over the 5 GHz unlicensed band is currently under study in 3GPP Release 18 and uses LBT procedures, as detailed in~\cite{Weerackody2025SLU}. This combination ensures coverage of both licensed mmWave bands and unlicensed Sub-6 GHz spectrum, essential for real-time SL scheduling under heterogeneous coexistence constraints.

Each transmission mode occupies a predefined bandwidth and experiences link-specific path loss and fading. mmWave links (26/28 GHz) suffer from high attenuation and blockage, while the 5 GHz unlicensed band is more susceptible to interference from nearby Wi-Fi devices.

\subsection{Channel Model}

Wireless channels are modeled using a Rician fading distribution, capturing both line-of-sight (LOS) and non-line-of-sight (NLOS) propagation conditions. The channel coefficient \( H \) is expressed as
\begin{equation}
H = \sqrt{\frac{K}{K+1}} H_{\text{LOS}} + \sqrt{\frac{1}{K+1}} H_{\text{NLOS}},
\end{equation}
where \( K \) is the Rician \( K \)-factor, representing the ratio between LOS and NLOS power components. The LOS component is modeled as
\( H_{\text{LOS}} = e^{-j 2\pi d_{3D} / \lambda} \), where \( d_{3D} \) is the 3D distance between the transmitter and receiver, and \( \lambda \) is the carrier wavelength. The NLOS component is modeled as a zero-mean circularly symmetric complex Gaussian random variable~\cite{TR38901}, with \( h_{\text{NLOS}} \) for mmWave bands (26 GHz and 28 GHz) following \( \mathcal{CN}(0, \frac{1}{L}) \), and for the 5 GHz band, \( \mathcal{CN}(0,1) \), reflecting a richer scattering environment at lower frequencies~\cite{TR38901}.

Large-scale path loss is modeled using the urban microcell (UMi) street canyon model specified in 3GPP TS 38.901~\cite{TR38901}. For distances smaller than the breakpoint distance \( d_{BP} \), the LOS path loss is
$PL_{\text{LOS, 1}} = 32.4 + 21 \log_{10}(d_{3D}) + 20 \log_{10}(f_c)$,
and for larger distances beyond the breakpoint,
$PL_{\text{LOS, 2}} = 32.4 + 40 \log_{10}(d_{3D}) + 20 \log_{10}(f_c) - [9.5 \log_{10}(d_{BP}^2) + (h_t - h_r)^2]$,
where \( h_t \) and \( h_r \) are the antenna heights of the transmitter and receiver, respectively, and \( d_{BP} = \frac{4 h_t h_r f_c}{c} \) is the breakpoint distance, $f_c$ represents the carrier frequency (Hz), and \( c = 3 \times 10^8 (m/s) \) as the speed of light. For NLOS, the effective path loss is the maximum of the LOS path loss and an empirically modeled NLOS path loss, which is
$PL_{NLOS} = 22.4 + 35.3 \log_{10}(d_{3D}) + 21.3 \log_{10}(f_c) - 0.3 (h_{{r}} - 1.5)$.
At the receiver, the achievable transmission rate \( R \) is calculated using Shannon’s capacity formula $R = B \log_2(1 + \text{SNR})$, where \( B \) is the allocated bandwidth, and \( \text{SNR} = \frac{P_t |H|^2}{N_0 B} \) is the received signal-to-noise ratio, with \( P_t \) as the transmission power and \( N_0 \) as the noise variance.

\subsection{Traffic and Queueing Model}

The traffic model follows the FTP standards defined in 3GPP~\cite{Release12,TR36889}, with packet arrivals modeled as a Poisson process at an average rate of \( \lambda \).
Each arriving packet represents a data session with a fixed size of 0.5 MBytes, typical of small file transfers in broadband wireless services. Upon arrival, packets are buffered and await transmission scheduling. The queue is modeled as a finite-capacity M/M/1 system, assuming exponential inter-arrival and service times, with overflow resulting in immediate blocking. While the gNB can select from five transmission modes (CC-28G, CC-26G, SL-L-28G, SL-L-26G, SL-U-5G), the underlying queuing model remains M/M/1, as only one packet is scheduled for transmission per decision epoch.
At each epoch, the DDQN agent observes the current queue status, channel conditions, and resource availability to select the optimal transmission mode and frequency band for the head-of-line (HOL) packet. For SL-U transmissions in the unlicensed band, coexistence with Wi-Fi is modeled using LBT and CSMA/CA procedures. In this context, the proposed agentic DDQN-based scheduler integrates these heterogeneous observations, enabling context-driven reasoning that balances blocking probability, throughput, and coexistence fairness across licensed and unlicensed bands.

\section{Problem Formulation}

The proposed framework departs from conventional RL by combining three design elements: (i) rich perception via a multi-dimensional state, (ii) goal alignment through a capacity-aware reward and throughput constraint, and (iii) a holistic action space across licensed and unlicensed modes. 
Together, these elements instantiate agentic behavior, where the agent perceives heterogeneous context, reasons about QoS trade-offs, and acts over all feasible mode–band options, rather than simply maximizing a scalar reward.

To enable adaptive mode and band selection in stochastic environments, we model the decision-making process as a discrete-time Markov decision process (MDP). At each decision epoch, the scheduling agent observes the system state, which includes the queue status, link quality indicators, and coexistence dynamics, and selects a transmission mode-band pair to maximize long-term performance. The objective is to balance throughput maximization with blocking probability control, leading to a constrained MDP (CMDP) formulation.

The discrete-time MDP is characterized by the tuple $(\mathcal{S}, \mathcal{A}, \mathcal{P}, \mathcal{R}, \gamma)$, where $\mathcal{S}$ represents the state space, which reflects the system status; $\mathcal{A}$ is the action space consisting of five transmission options; $\mathcal{P}$ is the state transition probability, which depends on scheduling actions and random traffic dynamics; $\mathcal{R}$ is the reward function that evaluates the immediate quality of scheduling decisions; and $\gamma \in (0,1]$ is the discount factor that controls the trade-off between immediate and future rewards.

\subsection{State Space}

At each decision epoch, the scheduling agent observes the following state features:
The normalized queue occupancy is defined as the ratio of the current queue length to the maximum capacity.
The residual resource ratios of five candidate bands (CC-28G, CC-26G, SL-L-28G, SL-L-26G, SL-U-5G), calculated as the current available bandwidth divided by the nominal full bandwidth.
The Wi-Fi idle probability indicates the likelihood that the unlicensed spectrum is not occupied by Wi-Fi transmissions.
These observations are concatenated into a seven-dimensional real-valued state vector provided as input to the learning agent,
enabling context-driven reasoning aligned with agentic AI principles.

\subsection{Action Space (Mode and Band Selection)}

The action space $\mathcal{A} = \{0,1,2,3,4\}$ consists of five transmission options, namely CC over the 28 GHz licensed band, CC over the 26 GHz licensed band, SL-L over the 28 GHz licensed band, SL-L over the 26 GHz licensed band, and SL-U over the 5 GHz unlicensed band, respectively. At each decision epoch, the agent selects one action from $\mathcal{A}$ to determine the transmission mode and frequency band for the HOL packet.

\subsection{Proposed Reward Function}

At each decision epoch, the agent receives the reward
\begin{equation}
r_t =
\begin{cases}
B\log_2(1+\text{SINR}_t), &\text{If transmission succeeds}, \\
0, &\text{If the packet is blocked}.
\end{cases}
\end{equation}
where \( B \) represents the bandwidth (in Hz) of the selected transmission band, and \( \text{SINR}_t \) is the received signal-to-interference-plus-noise ratio corresponding to the selected transmission mode and band.
This rate-based reward function incentivizes the agent to prioritize high-throughput, low-congestion links, aligning with agentic AI’s goal-oriented decision-making paradigm. 
In contrast to prior binary-reward schemes~\cite{Zou2019DRL, Pei2021Q}, this formulation incorporates link quality and bandwidth utilization directly into the reward, thereby encouraging QoS-aware decisions.
This capacity-aware reward, together with the long-term throughput constraint, aligns decisions with QoS objectives and discourages degenerate reward-chasing behaviors, thereby instantiating an agentic learning signal.

\subsection{State Transition Dynamics}

Given the current state \(s_t\) and action \(a_t\), the system transitions to a new state \(s_{t+1}\) based on stochastic dynamics, including packet arrivals (Poisson process), transmission outcomes (success or failure), channel fading variations, and Wi-Fi contention effects. These uncertainties necessitate adaptive scheduling to maintain performance under dynamic network conditions.

\subsection{Objective Function and Learning Formulation}

The objective of the proposed DDQN-based band allocation is to minimize the long-term average blocking probability while ensuring that the expected throughput exceeds a minimum service threshold. Formally, the optimization problem is formulated as
\begin{align}
\pi^* &= \arg\min_{\pi} \limsup_{T \to \infty} \frac{1}{T} \sum_{t=0}^{T-1} \mathbbold{1}_{\{B_t\}}, \tag{5a} \label{eq:obj_blocking} \\
\text{subject to} \quad &\liminf_{T \to \infty} \frac{1}{T} \sum_{t=0}^{T-1} \mathbb{E}_{\pi} \left[ R_t \right] \geq \epsilon_{R}, \tag{5b} \label{eq:throughput_constraint}
\end{align}
where \( \pi \) represents the scheduling policy learned by the DDQN agent, \( T \) is the time horizon over which performance is averaged, typically corresponding to the episode length or simulation window, \( B_t \) represents the event of packet blocking at time \( t \), and \( \mathbbold{1}_{\{B_t\}} \) is an indicator function that equals $1$ if the packet is blocked, and $0$ otherwise. \( R_t = B \log_2(1 + \text{SINR}_t) \) represents the instantaneous link capacity. \( \mathbb{E}_{\pi}[\cdot] \) represents the statistical expectation over trajectories generated by policy \( \pi \). \( \epsilon_R \) is the minimum acceptable long-term throughput requirement.
The goal in~(\ref{eq:obj_blocking}) is to reduce the blocking rate, particularly under constrained spectrum resources. The throughput constraint~(\ref{eq:throughput_constraint}) ensures that the agent does not minimize blocking by simply avoiding all traffic, thereby maintaining system utility.

Due to the complexity of wireless environments and the coexistence challenges posed by incumbent Wi-Fi systems, directly solving the CMDP becomes computationally intractable. As such, we adopt an agentic AI-driven DDQN framework to approximate the optimal action-value function \( Q^*(s, a) \) through agentic DDQN-based learning. Here, the Q-function estimates the expected cumulative discounted reward of taking action \( a \) in state \( s \), enabling the agent to make intelligent scheduling decisions. As the reward is shaped by achievable link capacity, the DDQN agent is naturally guided toward high-throughput decisions. Although the agent does not receive explicit blocking penalties, the resulting policy effectively minimizes blocking under spectrum constraints through rate-aware prioritization.

\section{Proposed Agentic DDQN Band Allocation}
Our agent retains the standard DDQN update rule and network, while its \emph{agentic} behavior arises from three elements: enriched perception via a multi-dimensional state, a QoS-aligned reward with constraint, and a unified mode–band action space. Building on these elements, the framework learns an optimal scheduling policy \( \pi^* \) through interactions with the environment, effectively balancing throughput and blocking under dynamic coexistence.

\subsection{Network Architecture}

The proposed DDQN framework employs two deep neural networks, an online network \(Q(s, a;\boldsymbol{\theta})\) and a target network \(Q(s ,a;\boldsymbol{\theta}^{-})\), sharing the same architecture. Here, \(\boldsymbol{\theta}\) and \(\boldsymbol{\theta}^{-}\) represent the trainable parameters of the online and target networks, respectively. The input is a real-valued state vector, and the output is a five-dimensional vector representing the Q-values of each transmission action.
Each network consists of an input layer, two fully connected hidden layers with 128 and 64 neurons activated by ReLU functions, and a linear output layer mapping to the discrete action set \(\mathcal{A}\). 

\subsection{Training Procedure}

At each decision epoch, the agent selects an action \(a_t\) using an \(\epsilon\)-greedy strategy: with probability \(\epsilon\), a random action is selected; otherwise, the action maximizing the Q-value from the online network is chosen.
The observed transition tuple \((s_t, a_t, r_t, s_{t+1})\) is stored in a fixed-size replay buffer \(\mathcal{D}\). Mini-batches sampled from \(\mathcal{D}\) are used to update the online network parameters \(\boldsymbol{\theta}\) by minimizing the Bellman loss
\begin{equation}
L(\boldsymbol{\theta}) = \mathbb{E} \left[ \left( y_t^{\text{DDQN}} - Q(s_t,a_t;\boldsymbol{\theta}) \right)^2 \right],
\label{eq:bellman_loss}
\vspace{-0.1in}
\end{equation}
where the target value \(y_t^{\text{DDQN}}\) is computed as
\begin{equation}
y_t^{\text{DDQN}} = r_t + \gamma Q\left( s_{t+1}, \arg\max_{a'} Q(s_{t+1}, a'; \boldsymbol{\theta}); \boldsymbol{\theta}^{-} \right).
\label{eq:target_value}
\end{equation}
Here, \(\boldsymbol{\theta}^{-}\) represents the parameters of the target network. Using separate networks mitigates overestimation bias by decoupling action selection and evaluation.
The overall DDQN-based scheduling and training process is summarized in Algorithm~\ref{alg:ddqn_band}.

\begin{algorithm}
\caption{Proposed Agentic DDQN Band Allocation}\label{alg:ddqn_band}
\SetKwInput{KwData}{Input}
\SetKwInput{KwResult}{Output}
\KwData{
Initial online network parameters \(\boldsymbol{\theta}\) and target network parameters \(\boldsymbol{\theta}^{-}\);\\
Replay buffer \(\mathcal{D}\); discount factor \(\gamma\);\\
Exploration parameters \((\epsilon_0, \epsilon_{\text{min}}, K_{\text{decay}})\);\\
Learning rate and optimizer settings;\\
Target network update frequency \(C\).
}
\KwResult{
Trained scheduling policy \(\pi(s) = \arg\max_a Q(s,a;\boldsymbol{\theta})\).
}
Initialize online network \(Q(s,a;\boldsymbol{\theta})\) and target network \(Q(s,a;\boldsymbol{\theta}^{-})\) with random weights\;
Initialize replay buffer \(\mathcal{D}\)\;
\For{each episode}{
    Initialize state \(s_0\)\;
    \For{each decision epoch \(t\)}{
        Update exploration rate: \(\epsilon_t = \max(\epsilon_{\text{min}}, \epsilon_0 \cdot e^{-t / K_{\text{decay}}})\)\;
        Generate a random number \(p \sim \mathcal{U}(0,1)\)\;
        \eIf{\(p < \epsilon_t\)}{
            Select a random action \(a_t\) from the action space\;
        }{
            Select action \(a_t = \arg\max_a Q(s_t, a; \boldsymbol{\theta})\)\;
        }
        Execute action \(a_t\), observe reward \(r_t\) and next state \(s_{t+1}\)\;
        Store transition \((s_t, a_t, r_t, s_{t+1})\) into \(\mathcal{D}\)\;
        Sample a mini-batch from \(\mathcal{D}\)\;
        Compute target values \(y_t^{\text{DDQN}}\) according to~(\ref{eq:target_value})\;
        Update online network parameters \(\boldsymbol{\theta}\) by minimizing the loss in~(\ref{eq:bellman_loss})\;
        \If{\(t \bmod C == 0\)}{
            Update target network: \(\boldsymbol{\theta}^{-} \leftarrow \boldsymbol{\theta}\)\;
        }
    }
}
\end{algorithm}

\vspace{-0.2in}
\subsection{Target Network Update}

The target network parameters \(\boldsymbol{\theta}^{-}\) are periodically updated to match the online network parameters \(\boldsymbol{\theta}\) every \(C\) decision epochs, i.e., \(\boldsymbol{\theta}^{-} \leftarrow \boldsymbol{\theta}\), providing a stationary target for stable learning.

\subsection{Exploration Strategy}

The exploration probability \(\epsilon\) is initialized at \(\epsilon_0\) and decays linearly to a minimum value \(\epsilon_{\text{min}}\) over \(K_{\text{decay}}\) training steps. Formally, at training step \(k\), it is updated as
\begin{equation}
\epsilon_k = \max\left(\epsilon_{\text{min}}, \epsilon_0 - \frac{\epsilon_0 - \epsilon_{\text{min}}}{K_{\text{decay}}} \times k\right).
\end{equation}
This schedule ensures sufficient early exploration while gradually shifting toward exploitation as learning progresses.

\section{Simulation Results}

We evaluate the proposed DDQN-based band allocation scheme in a heterogeneous wireless system comprising five candidate transmission modes: CC-28G, CC-26G, SL-L-28G, SL-L-26G, and SL-U-5G. Wi-Fi operates on the unlicensed 5 GHz band, and SL-U coexists with it using LBT via CSMA/CA. The gNB schedules one packet per epoch based on the current system state, as described in Section~\ref{section:system_model}.

We compare three scheduling policies: (i) the proposed DDQN agent, trained using a rate-aware reward function; (ii) a threshold-based heuristic adapted from~\cite{Chou2024ICC}, which disables SL-U access when the Wi-Fi idle probability drops below $50\%$, and prioritizes CC or SL-L based on fixed bandwidth thresholds; and (iii) a random selection baseline where each mode-band pair is chosen uniformly without regard to queue status, system load, or channel contention. This comparison isolates the benefit of adaptive learning from static or stochastic rule-based schemes.

To reflect real-world coexistence challenges, the unlicensed band is fixed at $100$ Mbps, emulating high Wi-Fi activity. The licensed spectrum budget varied from $500$ to $1000$ Mbps to observe blocking and throughput trends across different provisioning levels. Each mode occupies a predefined bandwidth, and channels are modeled using the 3GPP UMi-Street Canyon fading model with Rician fading ($K = 10$). The transmit-receive distances are set to $100$ m for CC links and $5$ m for SL and Wi-Fi nodes. The SNRs are configured as follows: $40$ dB for CC links, $13$ dB for SL-L, and $0$ dB for SL-U. The $0$ dB value for SL-U is aligned with 3GPP TR 38.889~\cite{TR38889}, which models NR-U transmissions under high contention in dense urban settings where unlicensed transmissions often suffer from interference and limited power. Similar near-zero SNR assumptions are adopted in~\cite{Pei2021Q, Bajracharya2023Bandit} to emulate practical coexistence with incumbent Wi-Fi deployments.

The DDQN agent is implemented as a fully connected neural network with two hidden layers of sizes $1024$ and $512$, using ReLU activations and a dropout rate of $0.3$. The output is a five-dimensional Q-value vector corresponding to the action space. Training is conducted with the Adam optimizer, an initial learning rate of $10^{-7}$ decayed to $10^{-9}$ in stages, and a batch size of $64$. The replay buffer holds $10^6$ transitions, and exploration follows an $\epsilon$-greedy strategy with $\epsilon$ linearly decaying from $1$ to $0.05$ at a rate of $5 \times 10^{-7}$. The model is trained for $3$ million episodes.

\begin{figure}[t]
\centering
\includegraphics[width=0.45\textwidth]{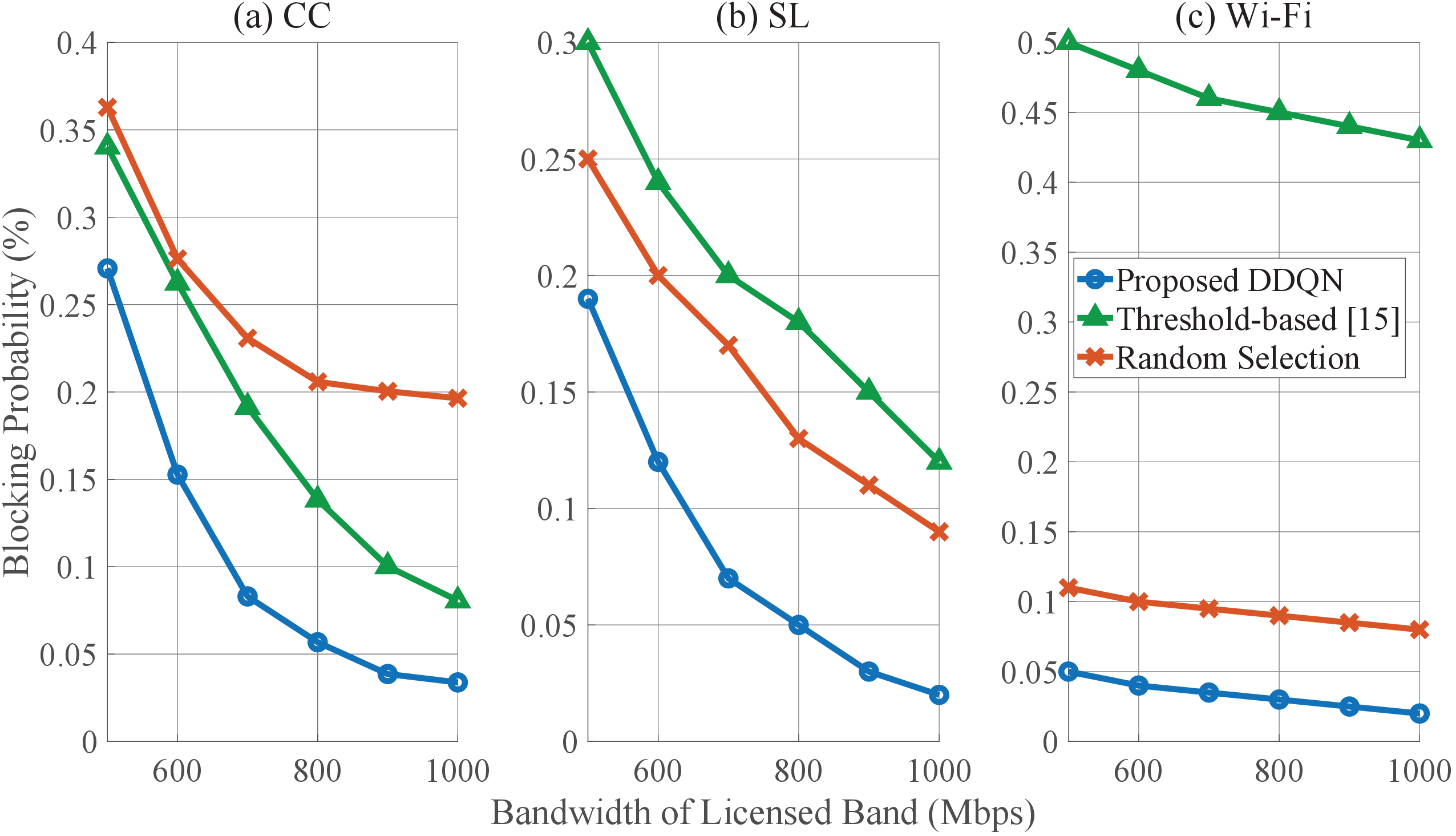}
\caption{Blocking probabilities over varying licensed bandwidths ($B_U = 100$ Mbps).}
\label{Results}
\vspace{-0.1in}
\end{figure}

Fig.~\ref{Results} illustrates the blocking probabilities of different methods as licensed bandwidth increases. The proposed DDQN outperforms both baselines across all bandwidth levels. For example, when the licensed bandwidth is constrained at $500$ Mbps, the blocking rate under DDQN is only $0.014$, compared to $0.112$ under the threshold-based policy and $0.08$ under random selection. This corresponds to a blocking rate reduction of approximately $87.5\%$ compared to the threshold-based method and $82.5\%$ compared to random selection. These results demonstrate that DDQN can effectively offload traffic to unlicensed bands when appropriate, while static thresholds often underutilize available spectrum, leading to congestion.

\begin{figure}[t]
\centering
\includegraphics[width=0.35\textwidth]{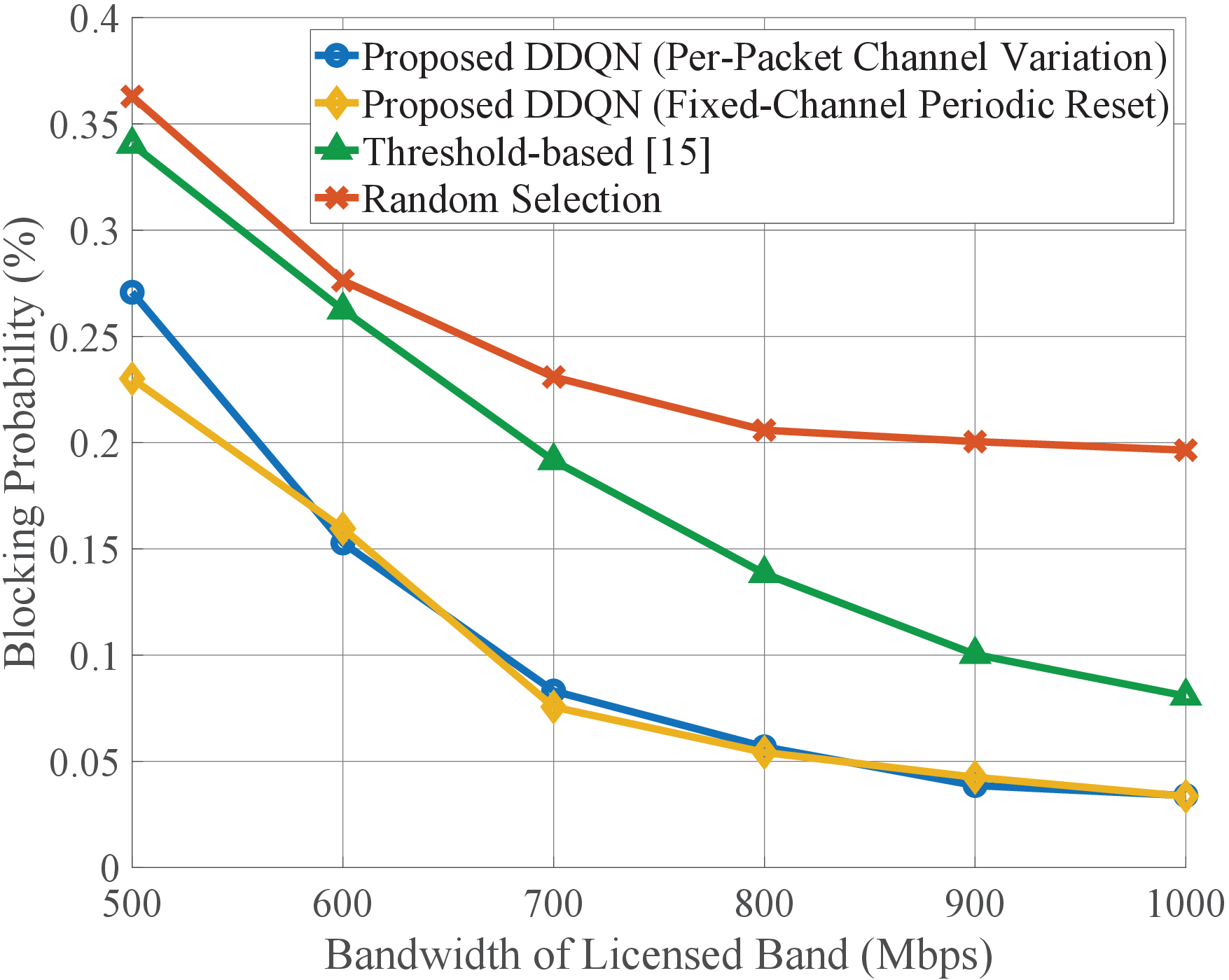}
\caption{Blocking probability of DDQN under static vs dynamic channel variation.}
\label{Results2}
\vspace{-0.2in}
\end{figure}

In Fig.~\ref{Results2}, we examine DDQN performance under two types of channel variation: (i) static channel per episode (i.e., channels remain fixed until buffer reset), and (ii) dynamic channel per packet (i.e., new channel realizations per transmission). The DDQN agent generalizes well in both settings, but achieves slightly lower blocking under per-packet variation due to its exposure to richer training samples. This confirms the agent's robustness and its ability to leverage short-term channel fluctuations for better resource scheduling.

\section{Conclusion}
In this paper, we propose an agentic DDQN-based scheduler for licensed and unlicensed band allocation in NR sidelink networks. Different from vanilla DDQN applications, the design couples rich perception, a goal-aligned reward, and a unified action space, enabling context-driven decisions. Simulations showed up to $87.5\%$ and $82.5\%$ blocking reduction versus threshold-based and random policies at $500$ Mbps licensed bandwidth, while maintaining throughput. These results highlight the promise of agentic DDQN as an E-agent for QoS-aware, adaptive, and stable scheduling in coexistence-limited NR SL networks, with future extensions toward H-agents that fuse embodied learning with foundation-model reasoning.

\end{document}